\documentclass[%
reprint,
superscriptaddress,
 amsmath,amssymb,
 aps,
prb,
showkeys
]{revtex4-2}

\usepackage{romannum}
\usepackage{booktabs}
\usepackage{graphicx}
\usepackage{lipsum}
\usepackage{placeins}
\usepackage{amsmath}
\usepackage{hyperref}
\hypersetup{
    colorlinks = true,
    allcolors = {blue},
}
\usepackage{xspace}
\newcommand{\lc}{\textsubscript}

\usepackage{epstopdf}

\newcommand{\andGang}{\textit{et al.}\xspace}    
\newcommand{\CFB}{Co\lc{40}Fe\lc{40}B\lc{20}\xspace}
\newcommand{\LTO}{LiTaO\lc{3}\xspace}
\newcommand{\Lam}{$\lambda/2$-plate\xspace}

\usepackage{graphicx}
\usepackage{dcolumn}
\usepackage{bm}

\begin{document}

\preprint{APS/123-QED}

\title{Phase-resolved imaging of coherent phonon-magnon coupling}

\author{Yannik Kunz} 
\email{ykunz@rptu.de}
\affiliation{Fachbereich Physik and Landesforschungszentrum OPTIMAS, Rheinland-Pfälzische Technische Universität Kaiserslautern-Landau, 67663 Kaiserslautern, Germany}
\author{Florian Kraft}
\affiliation{Fachbereich Physik and Landesforschungszentrum OPTIMAS, Rheinland-Pfälzische Technische Universität Kaiserslautern-Landau, 67663 Kaiserslautern, Germany}
\author{David Breitbach}
\affiliation{Fachbereich Physik and Landesforschungszentrum OPTIMAS, Rheinland-Pfälzische Technische Universität Kaiserslautern-Landau, 67663 Kaiserslautern, Germany}
\author{Kevin Künstle}
\affiliation{Fachbereich Physik and Landesforschungszentrum OPTIMAS, Rheinland-Pfälzische Technische Universität Kaiserslautern-Landau, 67663 Kaiserslautern, Germany}
\author{Torben Pfeifer}
\affiliation{Zernike Institute for Advanced Materials, University of Groningen, Groningen 9747 AG, The Netherlands}
\author{Matthias Küß}
\affiliation{Institute of Physics, University of Augsburg, 86135 Augsburg, Germany}
\author{Stephan Glamsch}
\affiliation{Institute of Physics, University of Augsburg, 86135 Augsburg, Germany}
\author{Manfred Albrecht}
\affiliation{Institute of Physics, University of Augsburg, 86135 Augsburg, Germany}
\author{Mathias Weiler}
\affiliation{Fachbereich Physik and Landesforschungszentrum OPTIMAS, Rheinland-Pfälzische Technische Universität Kaiserslautern-Landau, 67663 Kaiserslautern, Germany}

\date{\today}

\begin{abstract}
We use a direct phase-resolved optical technique to study the coherence of spin waves (SWs) that are driven by surface acoustic waves (SAWs) via resonant magnetoelastic coupling. For this, we employ a piezoelectric lithium tantalate (\LTO) substrate, equipped with micropatterned interdigital transducers for SAW excitation, which interact with SWs in a 5\;nm thin and 20\;µm wide \CFB-waveguide. We detect the SAW and the SW using a phase-locked micro-focused optical polarization detection experiment and use the characteristic polarization dependence to separate the SAW and SW signals. Our measurements directly image the resonant and coherent excitation of the SW by the SAW.
\end{abstract}

\keywords{Magnonics, Magnetoelastics, Phonon-Magnon-interactions}
\maketitle

\textit{Introduction.}
As nanoscale electronic circuits are reaching their physical limitations~\cite{IEEE_IRDS_2024_Metrology}, Spintronics provides a complementary approach for information transport and processing as spin currents convey the information by transportation of spin angular momentum~\cite{barlaSpintronicDevicesPromising2021, SpintronicsUltralowpowerCircuits2024, behin-aeinProposalAllspinLogic2010}.
This spin transport does not necessarily require a charge current~\cite{sander2017MagnetismRoadmap2017}, as the spin information can be transported by excitations of the magnetic order parameter, consequently bypassing the resistive energy loss to Joule heating~\cite{althammerPureSpinCurrents2018}. 
Spin currents can be generated, for instance, by the spin-Seebeck effect~\cite{uchidaObservationSpinSeebeck2008} or spin-pumping~\cite{brataasSpinBatteryOperated2002, tserkovnyakSpinPumpingMagnetization2002a, cornelissenLongdistanceTransportMagnon2015}.
The specialized field of magnonics~\cite{pirroAdvancesCoherentMagnonics2021} explicitly makes use of coherent spin currents in the form of spin waves (SWs). SWs with their quanta, the magnon, are the low energetic excitations of magnetically ordered systems, locally deflecting the magnetization from its equilibrium. The applications of coherent SWs range from signal processing~\cite{Mahmoud2021} to quantum computing~\cite{yuanQuantumMagnonicsWhen2022, Andrianov2014} and neuromorphic computing~\cite{Chumak2022, breitbach2025allmagnonicneuronsanalogartificial}. Especially for the latter, magnons are promising candidates due to their intrinsic non-linearity that is accessible at low powers~\cite{sergaYIGMagnonics2010, dreyerImagingPhaselockingNonlinear2022, breitbachNonlinearErasingPropagating2024}.\\
\begin{figure*}[t]
\includegraphics{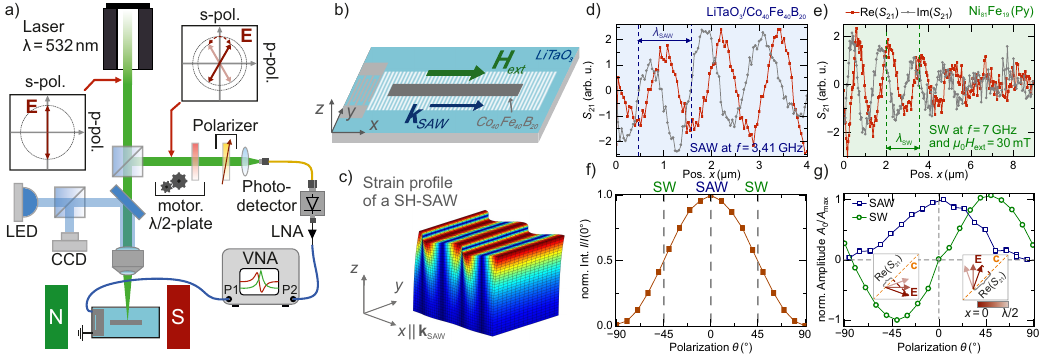}
\caption{\label{fig:Setup+Samp+Linescan+Pol}
(a) Schematic depiction of the experimental setup employing phase-locked micro-focused optical polarization analysis, commonly referred to as the \textit{µFR-MOKE}-technique. The reflected laser light is analyzed in its polarization and intensity state at fixed temporal wave phase $\omega t=\text{const.}$ (b) A sketch of the investigated sample: On a LiTaO\lc{3}-piezoelectric substrate, sets of IDTs excite coherent SH-SAWs toward a 5\;nm thick and 20\;µm wide \CFB-waveguide. (c) The in‑plane‑polarized shear‑strain profile of the propagating SH‑SAW mode. The strain amplitude of $\varepsilon_{xy}$ is shown in false color. (d) Example line-scan measurement of a SH-SAW mode on \LTO, measured at $f=3.41\;$GHz. The real and imaginary part of the $S_\text{21}$ parameter, measured by the VNA, encode the phase-information of the detected SAW. (e) Line-scan measurement of a SW on a 40\;nm thick permalloy film, excited by a microwave antenna, measured at $f=7\;$GHz and $\mu_0H_\text{ext}=30\;$mT in Damon-Eshbach geometry ($k_\text{SW}\perp\mu_0H_\text{ext}$). (f) The light intensity as a function of the polarization, analyzed by the rotatable $\lambda/2-$plate measured after the polarizer, relative to the incoming polarization on the sample, measured separately for SH-SAWs on \LTO and SWs on Py. The incoming, unmodulated light is linearly p-polarized. (g) Polarization dependence of the detected amplitude of the real-part of the SH-SAW and SW-signals, respectively. Notably, the SAW signal on \LTO maximizes at highest sensitivity on intensity modulation and not undergoing a sign change from negative to positive polarization analysis, showing an even symmetry. Comparably, the SW on Py is detected most efficiently at polarization modulating sensitivity and exhibiting a sign change from negative to positive polarization analyzing angle, representing an odd symmetry. The inset illustrates the origin of the sign flip of the SW detection, which occurs when the polarization is rotated due to a different selection of the analyzed polarization orientation.}
\end{figure*}
A key challenge for magnonics is that, although the SWs operate in a low-energy, low-dissipation regime, the coherent excitation by microwave antennas is rather inefficient, due to their weak coupling to microwave electronic cuircuits~\cite{kohl2025identificationminimizationlossesmicroscaled}. An approach to circumvent this is to exploit surface acoustic waves (SAWs), that couple via magnetoelastic interaction to the spin domain. The magnetoelastic mechanism for acoustic excitation of spin dynamics was already proposed by Kittel~\cite{kittelInteractionSpinWaves1958} and experimentally established shortly thereafter~\cite{bommelExcitationHypersonicWaves1959}. Previous studies showed that SAWs can be used to efficiently excite SWs linearly~\cite{kussGiantSurfaceAcoustic2023}, parametrically~\cite{janderParametricPumpingCoherent2025} and non-linearly~\cite{geilenParametricExcitationInstabilities2024}, even in ultra-low magnetic damping materials~\cite{Kunz2025, holzmannPolycrystallineYIGThin2025}, with the magnon-phonon interaction reaching the limit of strong coupling~\cite{kunstleMagnonpolaronControlSurface2025a, hwangStronglyCoupledSpin2024}.\\
A key necessity for magnonic devices that operate with SAW driven SWs is to retain a well-defined phase relation between the microwave, the SAW and the SW. For this, it is required to directly demonstrate that the SWs driven by the SAW are indeed coherent. Typical experiments employing electrical detection commonly only probe the absorption of the SAW in transmission experiments due to the SW excitation~\cite{kussSymmetryMagnetoelasticInteraction2021, weilerElasticallyDrivenFerromagnetic2011, weilerSpinPumpingCoherent2012}. These experiments do not provide evidence of the degree of coherence of the generated SWs, but only probe the SAW dissipation. Contrary, micro-optical experiments are a powerful tool for expanding research on phonon and magnon dynamics by providing space-, time-, and frequency-resolution~\cite{kraimiaTimeSpaceresolvedNonlinear2020, casalsGenerationImagingMagnetoacoustic2020, sebastianMicrofocusedBrillouinLight2015}. Previous studies using phase-resolved optical techniques were employed to identify resonant SAW-SW coupling and to develop methods of distinguishing individual signal contributions~\cite{kuszewskiOpticalProbingRayleigh2018a}. Our previous micro-focused Brillouin-light-scattering (µBLS) experiments provided evidence that the magnon-signal is coherent to the SAW signal by evaluating the interference of light scattered by SAWs and SWs~\cite{kunzCoherentSurfaceAcoustic2024}. In this work, we examine the direct phase relationship between the SAW and SW during resonant magnetoacoustic interaction, depending on space and the magnetic field.\\ 
Here, we present the direct imaging of resonant, coherently driven SWs by shear-horizontal SAWs (SH-SAWs). We achieve this by employing micro-focused frequency-resolved polar-magneto-optical Kerr-effect (MOKE) based spectroscopy, termed the \textit{µFR-MOKE} technique~\cite{liensbergerSpinWavePropagationMetallic2019}, an established technique alongside µBLS, that provides phase resolution similar to phase-resolved µBLS~\cite{sergaPhasesensitiveBrillouinLight2006, wojewodaPhaseresolvedOpticalCharacterization2023} or micro-focused super-Nyquist sampling MOKE~\cite{dreyerImagingPhaselockingNonlinear2022}. \textit{µFR-MOKE} is based on vector-network analysis  and can operate in a large frequency bandwidth with resolution down to 1\;Hz while providing a frequency range comparable to phase-resolved µBLS. \textit{µFR-MOKE} does not require the synchronization of pulsed lasers and electronics that is necessary for micro-focused super-Nyquist sampling MOKE.  We introduce a concept to separate SAW-generated signals from SW signature and show the intensity and phase modulation of the SAW. Finally, we directly demonstrate the resonant and coherent driving of SW precession by the SAW.

\textit{Experimental Methods}
To spatially resolve the magnetoelastic interaction, we use the experimental setup schematically shown in Fig.\;\ref{fig:Setup+Samp+Linescan+Pol}\;a), based on the~\textit{µFR-MOKE}-technique~\cite{liensbergerSpinWavePropagationMetallic2019}. Hereby, we use a linearly polarized $\lambda_\text{l}=532$\;nm continuous wave laser that is focused on the sample using a 100x/0.75\;NA microscope objective with a refraction limited spot size of approx. $d_\text{spot}=\lambda_\text{l}/(2\text{NA})\approx355$\;nm. The light interacts with the coherently excited acoustic or SW and is thereby modulated in polarization and intensity, depending on the probed spatial wave phase. The reflected light is coupled out using a beam splitter and passes a polarization analyzer unit consisting of a rotatable \Lam and a polarizer, converting the polarization modulation into an intensity modulation. This intensity modulation is captured by a fast broadband photodetector via an optical fiber. The photodetector's output passes through a low-noise amplifier and is connected to port 2 of a vector network analyzer (VNA). Port 1 of the VNA is used as a microwave output to excite the acoustic or SW on the sample. With the phase-resolving ability of the VNA we measure the complex-valued $S_{21}$-transmission parameter that is directly sensitive to the probed wave phase:
\begin{align}
    S_{21}=\frac{V_2}{V_1}&\propto m_\text{oop}\cdot\exp(ik_\text{SW}x);\hspace{0.2cm} \varepsilon_{ij}\cdot\exp(ik_\text{SAW}x),\\
    &S_{21}, m_\text{oop},\varepsilon_{ij}\in\mathbb{C}\notag.
\end{align}
Here, $V_{1,2}$ are the voltages at the output port $1$ and detection port $2$ of the VNA, $m_\text{oop}$ is the dynamic out-of-plane magnetization component $m_\text{oop}=M_\text{oop}/M_\text{S}$ and $\varepsilon_{ij}$ is the strain of the acoustic wave mode. A microscope picture and image stabilization is realized by a Köhler illumination and a CCD-camera that is coupled in and out of the light path by a dichroic mirror.\\
For SAW generation, we use piezoelectric \LTO as a substrate and sets of interdigital transducers (ITDs), as shown in Fig\;\ref{fig:Setup+Samp+Linescan+Pol}\;b). Details on the sample fabrication are given in the supplemental material~\cite{supplementalmaterial}. Importantly, the \LTO supports the excitation of a shear-horizontal SAW mode, which is an in-plane polarized shear wave with a dominant $\varepsilon_{xy}$ strain component that decays exponentially into the material along the $-z$-direction, as depicted in Fig.\;\ref{fig:Setup+Samp+Linescan+Pol}\;c). The result of an exemplary line scan measurement at a frequency of $f=3.41$\;GHz using a microwave power of $P_\text{MW}=12$\;dBm is shown in Fig.\;\ref{fig:Setup+Samp+Linescan+Pol}\;d), clearly demonstrating the spatially resolved wave fronts of the propagating SAW in $\text{Re}(S_{21})$ and $\text{Im}(S_{21})$. The detection mechanism relies on the elasto-optic effect, in literature also referred to as photoelastic, acoustooptic or Pockels-effect, and is explained by a modulation of the reflectivity and induced birefringence~\cite{thomsenSurfaceGenerationDetection1986, wempleTheoryElastoOpticEffect1970, wrightRealTimeImaging2005, matsudaReflectionTransmissionLight2002a, xieImagingGigahertzZerogroupvelocity2019}, which depends on the strain profile of the acoustic mode. The SAW can interact with the SWs in a 5\;nm thick and 20\;µm wide \CFB waveguide. Separately, we analyze the detection characteristics of SWs using a 40\;nm thick film of permalloy with a microwave antenna for direct SW excitation. A typical measurement result is depicted in Fig.\;\ref{fig:Setup+Samp+Linescan+Pol}\;e), revealing the SWs wavelength and decay length $\xi$ by its envelope. Here, we employed the Damon-Eshbach geometry, meaning the wave vector  $k_\text{SW}$ of the SW is perpendicular to the external magnetic field $\mu_0H_\text{ext}$ and excited at $f=7$\;GHz with $P_\text{MW}=0$\;dBm of microwave power. SWs are detected by the polar magneto-optical Kerr-effect that induces a circular dichroism and subsequent rotation of polarization of the reflected light on the analyzer plane~\cite{liensbergerSpinWavePropagationMetallic2019, hamrleAnalyticalExpressionMagnetooptical2010}.\\

\begin{figure}[t]
\includegraphics{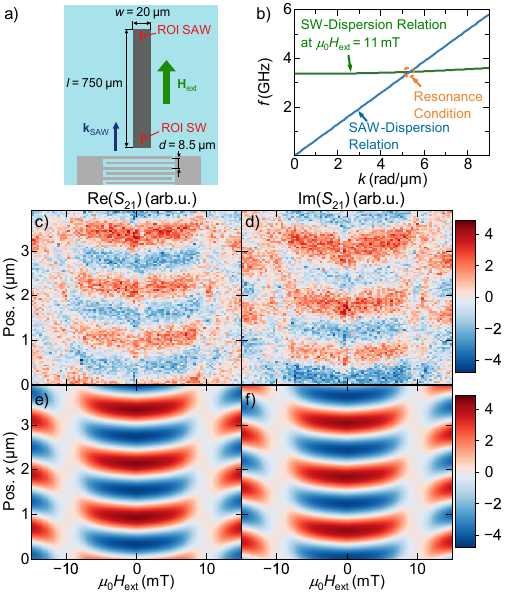}
\caption{\label{fig:SAW-fieldsweep} 
(a) Shows a sketch of the investigated structure and the measurement geometry. We orient the sample in backward-volume geometry ($k_\text{SAW}\|\mu_0H_\text{ext}$). The SAW-signal is picked up near the end of the waveguide (marked by ROI SAW). The SW signal shown in Fig.\;\ref{fig:SW-fieldsweep} is measured at the start of the \CFB-strip. (b) Shows the derived SAW dispersion relation (blue) and of the SW dispersion relation (green) at $\mu_0H_\text{ext}=11\;$mT computed using the thin-film approximation derived by Kalinikos-Slavin equation~\cite{kalinikosTheoryDipoleexchangeSpin1986}. The SAW frequency ($f=3.41$\;GHz) and wave vector ($k=5.3$\;rad/µm) excited by our IDT intersect the SW dispersion relation at the external field of $\mu_0H_\text{ext}=11\;$mT, fullfilling the resonance condition. (c-d) Show the measured SAW signal as $\text{Re}(S_{21})$ and $\text{Im}(S_{21})$ at $f=3.41$\;GHz, detected at "ROI SAW" indicated in (a) at 0° polarization analyzing position, when the SAW is detected most sensitively. At the resonance magnetic field of $\pm11\;$mT we observe a decrease in SAW amplitude as well as a phase shift, compared to the off-resonant case as a back action of the SW-system on the SAW. (e-f) Shows the fitted data for $\text{Re}(S_{21})$ and $\text{Im}(S_{21})$ using the theory by Dreher \textit{et al.}~\cite{Dreher2012} and Küß \textit{et al.}~\cite{kussNonreciprocalDzyaloshinskiiMoriya2020} using Eq.\;(11) given in the supplemental material~\cite{supplementalmaterial}.}
\end{figure}

\textit{Detection characteristics}
First, we focus on the properties of the detection mechanisms for SH-SAWs and SWs by our polarization analysis technique. For this, we rotate the analyzer's \Lam. As expected, the linear polarization exhibits a $\cos^2(\theta)$-dependence, where $\theta$ denotes the polarization angle relative to the pure p-polarized state of the incoming light, which is polarized parallel to the $\lambda/2$-plates neutral axis and the polarizer axis. This is depicated in Fig.\;\ref{fig:Setup+Samp+Linescan+Pol}\;f) and indicated by the normalized light intensity $I/I(0\text{°})$. We assess the obtained signal for both the SH-SAW and the SW under variation of the analyzed polarization angle, i.e. of the \Lam angle, relative to its neutral axis. For this, we measure a line scan as depicted in Figs.\;\ref{fig:Setup+Samp+Linescan+Pol}\;d) and e) and fit the complex $S_{21}$ data using~\cite{sebastianLowdampingSpinwavePropagation2012}:
\begin{align}
    S_{21}(x)=A_0\cdot e^{-\frac{x}{\xi}}\cdot e^{i(kx-\varphi_0)}.
\end{align}
Here, $A_0$ denotes the detected amplitude and $\varphi_0$ the phase. The result as a function of the analyzed polarization is shown in Fig.\;\ref{fig:Setup+Samp+Linescan+Pol}\;g) for the normalized detected amplitude $A_0/A_\text{max}$ extracted for the real-part. Remarkably, we find that the SH-SAW is detected most sensitively at an analyzer angle corresponding to maximized intensity modulation, see Fig.\;\ref{fig:Setup+Samp+Linescan+Pol}\;f) and is even in symmetry with respect to 0. Concluding, the detected signal dominantly stems from a reflectivity modulation induced by the SAW. 
Consequently, the incident electric field $\mathbf{E}_\text{in}=(E_\text{in},0,0)^T$ is modulated by the SAW phase according to $\mathbf{E}_\text{ref}=((r_{11}+\Delta r_{11}\cdot\exp(ikx))E_\text{in},0,0)^T$, with $r_{11}$ being the reflection coefficient and $\Delta r_{11}\cdot\exp(ikx)$ being the modulation that depends on the phase of the acoustic wave. Selection of the polarization orientation by the polarizer at an angle $\theta$ on the polarizer axis $\mathbf{c}=(\cos(\theta), \sin(\theta),0)^T$ and detection as a DC-voltage $V_2\propto[\mathbf{c}\cdot\mathbf{E}_\text{ref}]^2$ by the photo diode yields:
\begin{align}
    V_2\propto[(r_{11}+\Delta r_{11}\cdot\exp(ik_\text{SAW}x))E_\text{in}\cdot\cos(\theta)]^2.
\end{align}
This expression is equivalent to Eq.\;(1) Ref.\;~\cite{santosAcousticFieldMapping1999}. From this, it becomes clear that the maximum sensitivity to intensity modulation is obtained at a polarization of 0° with 180° periodicity due to the $\cos^2(\theta)$-dependence. Next, considering SW detection, we obtain the maximum signal at analyzer angles that maximize sensitivity to polarization modulation. This aligns with the expectation for the induced circular dichroism by the polar MOKE, which leads to a rotated polarization state on the analyzer plane, depending on the probed SW phase. Notably, we find that the detected real-part amplitude exhibits a sign change when moving from negative to positive analyzer angles, being odd in symmetry with respect to 0, as shown in Fig.\;\ref{fig:Setup+Samp+Linescan+Pol}\;g). 
To understand this, we consider the Kerr-rotated signal in simplified limit given by $\mathbf{E}_\text{ref}=(r_{11}E_\text{in}, \Theta_\text{K,0}\cdot\exp(ikx)r_{12}E_\text{in}, 0)^T$, with $\Theta_{\text{K,}0}\cdot\exp(ikx)$ denoting the polar Kerr angle and $r_{12}$ the off-diagonal component of the reflectivity tensor. After polarization selection and DC voltage detection (see above), we obtain:
\begin{widetext}
\begin{align}
    V_2&\propto[r_{11}E_\text{in}\cos(\theta) + \Theta_\text{K,0}\cdot\exp(ik_{\text{SW}}x)r_{12}E_\text{in}\cdot\sin(\theta)]^2\\
    &\approx r_{11}^2E_\text{in}^2\cos^2(\theta)+r_{11}r_{21}E_\text{in}^2\Theta_\text{K}\exp(ik_\text{SW}x)\sin(2\theta).
\end{align}
\end{widetext}
We hereby neglect the quadratic terms $\Theta^2_\text{K}\approx0$. As the first term contains only a constant background, not containing any SW information, we do not consider this part further. The second term includes the polarization modulation. Here $\text{Re}(V_2^2)$ and $\text{Im}(V_2^2)$ maximize for $\sin(2\theta)=\pm45$°. Importantly, as $\sin(2\theta)$ is odd in symmetry with respect to 0°, the detected phase flips by 180° when rotating from $-\theta$ to $+\theta$. This flip is depicted in the insets in Fig.\;\ref{fig:Setup+Samp+Linescan+Pol}\;g) by the different projection of the GHz-dynamic part of the electric field vector on the polarizer axis. At -45°, one obtains a minimum position after the selection of the polarization orientation on the polarizer, which is denoted by $\mathbf{c}$, while on +45°, one obtains a maximum.
The observed behavior is in contrast to the SAW and SW detection using mirco-focused Brillouin light scattering spectroscopy, where SAW and SW signals are in 90° polarization rotation to each other~\cite{geilenFullyResonantMagnetoelastic2022}. The polarization dependent footprints allow to discriminate SAW and SW signals. By subtracting the measured $\text{Re}(S_{21})$ and $\text{Im}(S_{21})$ at -45° and +45°, the SAW signal is effectively diminished, while the SW signal is preserved by exploiting the odd and even symmetry correlations. 
The corresponding angles are marked in Fig.\;\ref{fig:Setup+Samp+Linescan+Pol}\;f).

\textit{Phase-resolved SAW-SW-coupling}
Now, we focus on the phase-resolved detection of magnetoelastic SW excitation by SAWs. For this, we place our \LTO/\CFB sample, oriented with $k_\text{SAW}$ parallel to the external magnetic field $\mu_0H_\text{ext}$, commonly called the backward-volume geometry. Thereby, the shear strain component generates an effective magnetoelastic field $\mu_0h_\text{ip}$ given by:
\begin{align}
    \mu_0h_\text{ip}=-\frac{2b_1\varepsilon_{xy}}{M_\text{S}}\cos[2\sphericalangle(\mathbf{k}_\text{SAW},\mathbf{M})],
\end{align}
where $b_1$ is the magnetoelastic coupling constant of \CFB and $\mathbf{M}$ is the magnetization~\cite{Dreher2012}. We choose this configuration, as the SH-SAW $\varepsilon_{xy}$-component couples efficiently in the orientation $\textbf{k}_\text{SAW}||\textbf{M}||\mathbf{H}_\text{ext}$, as shown by Küß \andGang~\cite{kussSymmetryMagnetoelasticInteraction2021, kussChiralMagnetoacoustics2022a}. We excite the SH-SAW at $f=3.41$\;GHz with $P_\text{MW}=12$\;dBm and first focus on the end of the \CFB-waveguide, marked by "ROI SAW" in Fig.\;\ref{fig:SAW-fieldsweep}\;a). As shown in Ref.~\cite{kunzCoherentSurfaceAcoustic2024}, the total SAW absorption is greatest at the furthest point from the exciting IDT due to the exponential decay of the acoustic wave as it propagates.\\
We expect resonant SAW-SW coupling when the excited SAWs frequency and wave vector on the SH-SAW dispersion relation match the SW dispersion relation. This is met at an externally applied magnetic field of $\mu_0H_\text{ext}=\pm11$\;mT, as shown in Fig.\;\ref{fig:SAW-fieldsweep}\;b). During the experiment, we rotate the \Lam to $\theta=0$° analyzing position on maximized SAW detection efficiency, and measure linescans for each applied magnetic field. The result is shown in Figs.\;\ref{fig:SAW-fieldsweep}\;c) and d) for the obtained $\text{Re}(S_{21})$ and $\text{Im}(S_{21})$. As is evident, at the field of $\pm11$\;mT the SAW amplitude is strongly decreased and the phase of the acoustic wave is shifted, as one sweeps over the resonance magnetic field. This occurs, as when fulfilling the resonance condition, the acoustic impedance of the material is altered. Consequently, the acoustic absorption and group velocity are modulated, which yields the observed phase shift. This is described by the theory developed by Dreher \andGang~\cite{Dreher2012} and Küß \andGang~\cite{kussNonreciprocalDzyaloshinskiiMoriya2020} by:
\begin{align}
    \varepsilon_{xy}(x)&\propto\exp\left(-\frac{1}{2}\text{Im}(\mathbf{h^*}\bar\chi\mathbf{h})x\right),\\
    \Delta\phi(x)&\propto\text{Re}(\mathbf{h^*}\bar\chi\mathbf{h})x.\label{eq:DeltaPhi}
\end{align}
\begin{figure}[t]
\includegraphics{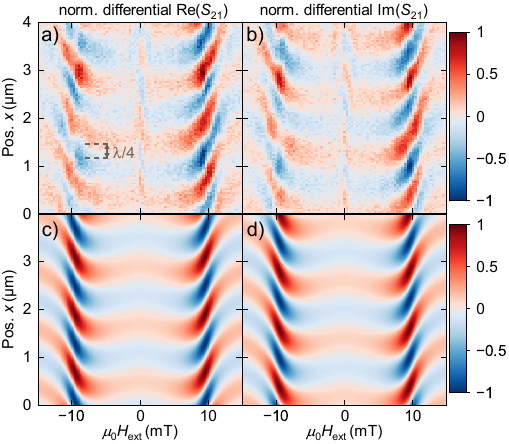}
\caption{\label{fig:SW-fieldsweep}
Spin-wave signal extracted in the ROI SW region, indicated in Fig.\;\ref{fig:SAW-fieldsweep}\;a) near the start of the ferromagnetic waveguide. At an external magnetic field of $\mu_0H_\text{ext}=11\;$mT, the SAW resonantly drives the SW. This becomes apparent by the increase in the detected amplitude and the phase difference of 90° of the signal, as marked by the gray dashed lines. The result clearly shows that the SAW drives a coherent SW precession. (a-b) Show the experimental data, in (c-d) the fit according to Eq.\;(13) in the supplemental material~\cite{supplementalmaterial} is shown.}
\end{figure}
Hereby, $\varepsilon_{xy}(x)$ is the strain and $\Delta\phi(x)$ the associated phase shift along the propagation direction, $\mathbf{h}$ is the generated magnetoelastic driving field, $\bar\chi$ is the dynamical Polder susceptibility tensor, and $x$ denotes the wave's propagated distance over the \CFB. The expression for $\Delta\phi$ directly follows from the Kramers-Kronig-relations~\cite{Kramers1927, kronigTheoryDispersionXRays1926}.\\
Details on the modeling and the formulation of the fitted equations can be found in the supplemental material~\cite{supplementalmaterial}. Additionally, in Figs.\;\ref{fig:SAW-fieldsweep}\;c) and d) a slight increase in the obtained SAW amplitude with propagation length is visible. This is attributed to a slight deterioration in the reflected light coupled into the optical fiber caused by the stage moving over several micrometer distances, which improves again during small-step scanning of the ROI.\\
Finally, we move the detection ROI to the start of the \CFB layer as indicated by ROI SW in Fig.\;\ref{fig:SAW-fieldsweep}\;a). As previously described, we measure line-scans as a function of the applied magnetic field. To maximize the obtained response, we repeat the measurement at -45° and +45°, and subtract the absolute values of the resulting signal to suppress the SAW contribution and enhance the SW signature. The extracted normalized signal is shown in Figs.\;\ref{fig:SW-fieldsweep} a) and b) for $\text{Re}(S_{21})$ and $\text{Im}(S_{21})$, respectively.\\
As is evident in Fig.\;\ref{fig:SW-fieldsweep} by the strongly increased signal intensity at $\pm11$\;mT, the resonantly driven SW is detected alongside a 90° positional phase difference of the SW relative to the off-resonantly detected SAW signal. This is analogous to the characteristic 90° phase difference of a driven harmonic oscillator at resonance between the driving force, that is the SAW, and the resonantly driven system, that is the SW at the resonance magnetic field. As is visible when comparing Figs.\;\ref{fig:SAW-fieldsweep}\;c) and d) with Figs.\;\ref{fig:SW-fieldsweep}\;a) and b), the driven SW possesses the same wavelength $\lambda$ as the SAW. Profoundly, this measurement shows not only the resonant excitation of the SW but importantly directly demonstrates the coherent driving of the spin-waves precession. We note that the SW decay length in the magnetic structure is only about $\xi_\text{SW}\approx1.85$\;µm (SW group velocity $v_\text{G}\approx22.4$\;m/s~\cite{supplementalmaterial}), so continuous spatial driving by the SAW is crucial for the SW driving which we observe. The SW does not propagate freely, but rather represents resonant SW precession on the SW dispersion, driven by the SAW. Further noticeable is that off-resonantly, the SAW signal does not fully vanish. This is due to the shear strain of the SH-SAW inducing an optical birefringence, which gives rise to a polarization rotation. Similar to the measured SAW in Figs.\;\ref{fig:SAW-fieldsweep}\;c) and d) a small increase in obtained amplitude is visible in Figs.\;\ref{fig:SW-fieldsweep}\;a) and b), which we again denote to a small improvement of the coupling into the optical fiber during the scanning.\\
We employ the theoretical formalism by Dreher \textit{et al.}~\cite{Dreher2012} and Küß \textit{et al.}~\cite{kussNonreciprocalDzyaloshinskiiMoriya2020}, where we considered not only the resonantly driven spin-wave, but additionally a small signal contribution by the acoustic wave that is 90°-phase shifted to the SW signal. The result is shown in Figs.\;\ref{fig:SW-fieldsweep}\;c) and d). Details on the formulation can be found in the supplemental material~\cite{supplementalmaterial}. We obtain a good agreement between the experimental data and the theoretical modeling. Conclusively, the presented data evaluation can be used to extract the magnetic parameters, among them the spin-wave damping and decay rate, the magnetoelastic coupling parameters. By increasing the length of the scanning region, the SAW decay length and loss rate can directly be measured, too. This allows to estimate the coupling strength and the cooperativity of the system. Hereby, we first estimate the SAW and SW loss rate as $\kappa_\text{SAW}=12.0\pm1.2$\;MHz, following Eq.\;(16) and $\kappa_\text{SW}=1.41\pm0.2$\;GHz, to Eq.\;(14) in the supplemental material~\cite{supplementalmaterial}. We note, that in the ultra-thin film of 5\;nm, the effective magnetic damping is significantly enhanced by out-of-plane anisotropy effects and two magnon scattering processes, leading to increase of the SW loss rate. The coupling strength is then estimated to $g=6.63\pm0.9$\;MHz (Eq.\;(18)) and the corresponding cooperativity $C=g^2/(\kappa_\text{SAW}\kappa_\text{SW})=(2.6\pm0.9)\cdot10^{-3}$. Thereby, the device operates in the weak coupling limit, which could be enhanced, for instance, by using thicker \CFB-films or Bragg-gratings to create a planar cavity resonator~\cite{kunstleMagnonpolaronControlSurface2025a, hwangStronglyCoupledSpin2024}.

\textit{Summary.}
Our measurements demonstrate that, besides SWs, one can detect SAWs using a polarization-intensity analysis experiment. We presented a mean to effectively separate SAW and SW signatures by their distinct polarization footprints. This enables measuring the resonant magnetoelastic coupling between SAWs and SWs, and allows us to directly image the SAW attenuation and phase shift due to the modified acoustic impedance by the SW. Remarkably, we directly showed the resonant and coherent driving of SWs by SAWs, paving the way for the exploitation of magnetoelastic based SW excitation for coherent magnonic applications.

\textit{Acknowledgments.}
This work was supported by the European Research Council (ERC) under the European Union’s Horizon Europe research, innovation programme (Consolidator Grant ``MAWiCS", Grant Agreement No. 101044526) and by the Deutsche Forschungsgemeinschaft (DFG, German Research Foundation) by Project No. 492421737 and TRR 173—268565370 (project B01 and B13).

\textit{Data Availability.}
The data that support the findings of this article are openly available \cite{Zenodo}.

\nocite{*}


\bibliography{apssamp}

\end{document}